**Title**：Surface-wave pulse routing around sharp corners

**Author**：*Zhen Gao, Hongyi Xu, Fei Gao, Youming Zhang, Yu Luo\*, and Baile Zhang\**

Dr. Z. Gao, Dr. H. Xu, Dr. F. Gao, Dr. Y. Zhang, Prof. B. Zhang

Division of Physics and Applied Physics, School of Physical and Mathematical

Sciences, Nanyang Technological University, Singapore 637371, Singapore.

E-mail: blzhang@ntu.edu.sg (B. Zhang)

Prof. Y. Luo

School of Electrical and Electronic Engineering, Nanyang Technological University,

Nanyang Avenue, Singapore 639798, Singapore

E-mail: luoyu@ntu.edu.sg (Y. Luo)

Prof. B. Zhang

Centre for Disruptive Photonic Technologies, Nanyang Technological University,

Singapore 637371, Singapore.

**Abstract:** Surface plasmon polaritons (SPPs), or localized electromagnetic surface waves propagating on a metal/dielectric interface, are deemed promising information carriers for future subwavelength terahertz and optical photonic circuitry. However, surface waves fundamentally suffer from scattering loss when encountering sharp corners in routing and interconnection of photonic signals. Previous approaches enabling scattering-free surface-wave guidance around sharp corners are limited to either volumetric waveguide environments or extremely narrow bandwidth, being unable to guide a surface-wave pulse (SPP wavepacket) on an on-chip platform. Here, in a surface-wave band-gap crystal implemented on a single metal surface, we demonstrate in time domain routing a surface-wave pulse around multiple sharp corners without perceptible scattering. Our work not only offers a solution to on-chip surface-wave pulse routing along an arbitrary path, but also provides spatio-temporal information of the interplay between surface-wave pulses and sharp corners, both of which are desirable in developing high-performance, large-scale integrated photonic circuits.

# 1. Introduction

Surface plasmon polaritons (SPPs), generally described as electromagnetic (EM) surface waves on a metal/dielectric interface,[1] show a promising potential to merge photonics and electronics on the same metal circuitry,[2,3] because of their inherent subwavelength confinement. Although metals are remarkably lossy in the visible spectrum and beyond, they can be treated as perfect electric conductors (PECs) with negligible loss in the far-infrared, terahertz, and microwave spectra. Spoof SPPs are EM surface modes at these frequency bands, supported on a periodically textured metal surface.[4,5] They show great potential in building high-efficiency subwavelength photonic circuits and other unprecedented functional devices from microwave to infrared spectra.[6-12] However, while being routine in electronic circuits, near-perfect transmission through sharp corners is fundamentally difficult for (spoof) SPP waveguides because of the intrinsic scattering loss of guiding EM surface waves around sharp corners. The scattered energy will not only deteriorate transmission, but also introduce unwanted crosstalk of signals,[13] being especially detrimental in compactly integrated photonic circuits.

To date, scattering-free guidance of EM surface waves around sharp corners has been realized only in photonic topological insulators,[14-18] which support topological EM surface states as an analogue of electronic chiral edge states in topological insulators, and later in surface-wave transformation cloaks,[19] which apply transformation-optics to effectively warp the EM space around sharp corners analogously to the way in which gravity curves space in general relativity. However, these approaches generally require a volumetric waveguide environment (usually between two parallel metal plates) to achieve the out-of-plane confinement, or only apply to a very narrow bandwidth, being unable to manipulate surface-wave pulses (SPP wavepackets) on a single metal surface.

Nevertheless, the capability of manipulating surface-wave pulses is extremely desirable in ultrafast optical data processing and recording.[20-23] It is therefore necessary to have a clearer and more complete understanding in the spatio-temporal interplay between surface-wave pulses and the underlying photonic structures carrying them. The surface-wave pulse propagation at optical frequencies has been directly imaged along a straight path using the femtosecond laser pulse tracking technique. [24-28] However, how a surface-wave pulse interacts with a sharp corner remains unobserved in reality, not to mention the imaging of the process of routing a surface-wave pulse around sharp corners with near-perfect transmission.

It is the purpose of this article to develop an approach of routing surface-wave pulses around multiple sharp corners, especially with the spatio-temporal information in the routing process. We first show that by introducing a line defect in a photonic band-gap crystal for surface waves, the surface waves can be tightly guided along the line defect in the surface-wave band gap.[29-30] Then by tracking and mapping a surface-wave pulse as it propagates around multiple sharp corners in a surface-wave band-gap crystal, as schematically shown in **Figure 1**, we demonstrate that near-perfect transmission around multiple sharp corners can be achieved for surface-wave pulses.

The remarkable high-efficiency time-domain surface-wave pulse routing around multiple sharp corners in the surface-wave band-gap crystals is mainly because of the following two reasons. First, the wide forbidden band gap of the surface-wave band-gap crystal suppresses scattering of surface waves into the ambient space in a broad bandwidth. This feature is similar to previous wavelength-scale photonic circuits in a photonic crystal.[31-33] Second, because the eigenmode supported by the line-defect waveguide of the surface-wave band-gap crystal is essentially coupled toroidal resonance mode and electric quadrupole mode with four-fold rotational symmetry[34-36] (more details to be discussed later), the momentum of the surface EM waves before and after passing the sharp corners in an extremely compact space is inherently matched.[37] Thus there is no need to purposely construct extra local resonances[38,39] at the sharp corners in order to facilitate tunneling the EM energy through, being in sharp contrast to previous photonic crystal-based proposals.[31-33]

## 2. Experimental demonstration of the surface-wave band gap

We start the experimental demonstration with a surface-wave band-gap crystal that consists of a periodic (period $d = 5.0$ mm) $25 \times 25$ square lattice of circular metal rods with height $H = 5.0$ mm and radius $r = 1.25$ mm, all standing on a flat metallic surface, as shown in **Figure 2**a. The calculated band structure, as shown in **Figure 2**b, reveals a complete forbidden band gap for surface waves from 12.6 GHz to 27 GHz for the current surface-wave band-gap crystal. Using a vector network analyser (Agilent Technologies N5225A) and two homemade monopole antennas placed at two opposite sides of the experimental sample, we measured the continuous wave (CW) transmission spectrum supported by the surface-wave band-gap crystal, as shown in **Figure 2**c. The measured transmission exhibits a substantial drop from 12.6 GHz to 27 GHz, corresponding to the forbidden band gap of the surface-wave band-gap crystal.

## 3. Time-domain surface-wave pulse propagation along a straight line-defect waveguide

A straight line defect can be constructed by shortening the height of a row of rods in the middle from $H = 5.0$ mm to $h = 4.3$ mm, as indicated in **Figure 3**a. **Figure 3**b shows the dispersion (red line) of the line-defect waveguide that is calculated with the eigenmode solver of the commercial software CST Microwave Studio (See more details in Method in Supporting Information). Note that only the part below the light line and within the band gap (purple region) can confine and guide surface waves along the line-defect waveguide.

In fact, even without the surrounding surface-wave band-gap crystal, this row of shortened rods themselves can also guide surface waves. This is a typical traditional surface-wave waveguide (widely referred to as "domino plasmons"[8]) that has been studied extensively with various device applications.[10,40-46] For comparison, we also simulated the dispersion (blue line in **Figure 3**b) of surface waves guided along this domino-plasmon waveguide, when all the other rods of the surrounding surface-wave band-gap crystal were removed. It can be seen that this domino-plasmon waveguide exhibits a polaritonic dispersion that starts from the light line and then tends to zero group velocity at the Brillion zone edge.

We used an arbitrary waveform generator (Tektronix AWG70001A) to generate an 800 MHz wide Gaussian pulse centred around 13.4 GHz within the line-defect waveguiding band and launch it into the line-defect waveguide through an input monopole antenna. Using the other output monopole antenna connected to an oscilloscope (Tektronix MSO 72004C), we measured the pulse signal at the third rod from the input as the input pulse (grey line in **Figure 3**c; delay-time 0 ns) to minimize the interference from the source. Similarly, the output pulse (red line in **Figure 3**c) was measured at the third rod from the output. As shown in **Figure 3**c, it takes 6.2 ns for the pulse to propagate through the 100 mm-long straight line-defect waveguide.

**Figure 3**d shows the spectral signature of measured pulse signal. For comparison, the ideal pulse profile as the input to the arbitrary waveform generator is also shown. As the surface-wave pulse propagates along the straight line-defect waveguide, it maintains its amplitude and spectral signature without perceptible distortion.

In the following, we reconstruct the dynamic images of pulse propagation based on the microwave near-field imaging measurement. Two steps are adopted in the experiment. First, we mapped the surface-wave near-field distributions (electric field component $E_z$) on a transvers plane 1 mm above the line-defect waveguide, covering the frequency range from 13 GHz to 13.8 GHz with a step of 0.01 GHz (The spectral measurement and near-field imaging results are shown in Figure S1 in Supporting Information). Second, we reconstruct the original surface-wave pulse by using the measured surface-wave near-field distributions with different frequencies (see more details in Method in Supporting Information).

**Figure 3**e shows the reconstructed images of surface-wave pulse propagation through the line-defect waveguide. The pulse signal reaches the last rod at 6.4 ns, being consistent with the measured time-delay of 6.2 ns in **Figure 3**c. Movie S1 shows more details of the propagating pulse along the straight line-defect waveguide.

## 4. Time-domain surface-wave pulse propagation along a U-shaped line defect waveguide

We then examine the performance of a U-shaped line-defect waveguide with two right-angle sharp corners (**Figure 4**a). Using the same experimental method as in **Figure 3**, we measured the input pulse and the output pulse, which are shown in **Figure 4**b. The measured two pulse signals are almost the same after the pulse has propagated through two sharp corners.

We reconstruct the images of dynamic pulse propagation from the measured near-field distributions (The spectral measurement and near-field imaging results are shown in Figure S2). The reconstructed images are shown in **Figure 4**c. It can be seen that the pulse reaches the first and second sharp corners at 4 ns and 6 ns, respectively. The pulse is guided around both corners, leaving the U-shaped line-defect waveguide at 8 ns with imperceptible scattering. Movie S2 displays the whole process of the pulse propagation through the U-shaped line-defect waveguide.

For comparison, we also reconstruct the surface-wave pulse propagation along a U-shaped domino-plasmon waveguide, when all the other rods in the surrounding surface-wave band-gap crystal were removed. As presented in **Figure 4**d, the pulse reaches the first sharp corner at 1.4 ns and the second sharp corner at 2 ns. Both corners cause strong scattering and reflection, leading to almost zero transmission at the output.

Movies S3 shows more details about the process of the pulse through the U-shaped domino-plasmon waveguide.

**5. Mechanism of surface-wave pulse guidance around sharp corners**

To understand the underlying mechanism of the high-efficiency surface-wave pulse guidance around a sharp corner, we first simulate the magnetic and electric field distributions of an isolated point defect that can be realized by shortening only one single rod (black circle) in the surface-wave band-gap crystal, as shown in **Figure 5**a and **Figure 5**d, in which a transverse plane at the half height of the 5-mm-high metallic rods is adopted. The closed loop of magnetic field and the quadrupolar profile of electric field reveal that the eigenmode of a point defect in the surface-wave band-gap crystal is a localized magnetic toroidal mode and an electric quadrupole mode[34-36] with four-fold rotational symmetry.

Then we simulate the magnetic and electric field distributions of a line-defect waveguide with a sharp corner, as shown in **Figure 5**b and **Figure 5**e. We choose the frequency of 13.09 GHz for illustration in order to make the phase difference between fields of neighbouring shortened rods exactly equal to $\pi/2$, such that when the magnetic and electric fields are maximum around one shortened rod, they are zero around the neighbouring shortened rod. Thus the overlap of fields between neighbouring shortened rods will not obscure the mode profile of each shortened rod to be observed. It can be seen that, even in an array of closely positioned shortened rods, the magnetic toroidal mode and the electric quadrupole mode of each shortened rod still maintain their mode profiles with four-fold rotational symmetry. This is in sharp contrast to typical plasmonic waveguides formed by an array of plasmonic resonators.[47] For typical coupled plasmonic resonators, only when their coupling strength is weak can each resonator maintain its mode profile.[37] Yet in a surface-wave band-gap crystal, the four-fold rotational symmetry of mode profile around a shortened rod is maintained by the crystal symmetry, rather than the coupling strength between neighbouring shortened rods.

For comparison, we also plot the magnetic and electric field distributions of a domino-plasmon sharp bend, when all other rods in the surface-wave band-gap crystal are removed, as shown in **Figure 5**c and **Figure 5**f. The frequency of 12.00 GHz is used to set the phase difference between neighbouring rods equal to $\pi/2$. It can be observed that, without the surrounding surface-wave band-gap crystal, both the magnetic and the electric fields do not exhibit four-fold rotational symmetry.

The four-fold rotational symmetry of mode profile is the key to realize near-perfect guidance around a sharp corner. When propagating through the sharp bend, because of the presence of the preserved four-fold rotational symmetry, the surface waves feel as if they were still propagating along a straight waveguide. This is very similar to the previous coupled-resonator optical waveguides (CROW)[37] in which the rotational symmetry of each resonator causes the perfect transmission through a sharp bend, given that the mode profile of each resonator should not be perturbed by near-field coupling. On the other hand, without the four-fold rotational symmetry, serious scatterings occurs at the sharp corner in a typical domino-plasmon waveguide (**Figures. 5**c and **5**f).

## 6. Time-domain surface-wave pulse propagation around multiple sharp corners

Because the height of rods in the surface-wave band-gap crystal can be adjusted conveniently, it is feasible to form an arbitrary path for routing surface-wave pulses. In the following, we constructed a tortuous line-defect waveguide with multiple sharp corners (**Figure 6**a) and launched the same surface-wave pulse (central frequency 13.4 GHz and band width 0.8 GHz) to examine its performance. We measured the input pulse (grey line) and the output pulse (red line), as shown in **Figure 6**b. A time delay of 10.5 ns and only <10% amplitude decay can be observed after propagating through six sharp corners in the surface-wave band-gap crystal. To gain deeper insight of the dynamic evolution of the pulse in the tortuous waveguide, we reconstruct the images of the pulse at different time frames, as shown in **Figure 6**c. It is clear that the pulse signal can smoothly propagate around all six sharp corners without perceptible reflection or scattering, confirming the high transmission of surface waves through multiple sharp corners along an arbitrary path in the surface-wave band-gap crystal. Movies S4 shows more details about the dynamic evolution process of the pulse through tortuous line-defect waveguide with multiple sharp corners.

## 7. Conclusions and Discussion

In summary, by implementing a surface-wave band-gap crystal on a single metal surface, we provide a solution to routing surface-wave pulses around sharp corners, as a step forward towards solving a long-lasting bottleneck in highly integrated subwavelength photonic circuits. We applied a time-resolved near-field imaging technique to reveal the dynamics of a surface-wave pulse in both space and time as the pulse propagates through multiple sharp corners along an arbitrary path without perceptible scattering. The concept and method are general and can be directly applied to the terahertz and far-infrared frequency ranges to realize ultra-compact terahertz and photonic circuits with high-performance and high integration density.

## Supporting Information

Supporting Information is available from the Wiley Online Library or from the author.

## Acknowledgements

This work was sponsored by the NTU Start-Up Grants, Singapore Ministry of Education under Grant No. MOE2015-T2-1-070, MOE2011-T3-1-005.

Received:
Revised:
Published online:

# Figures

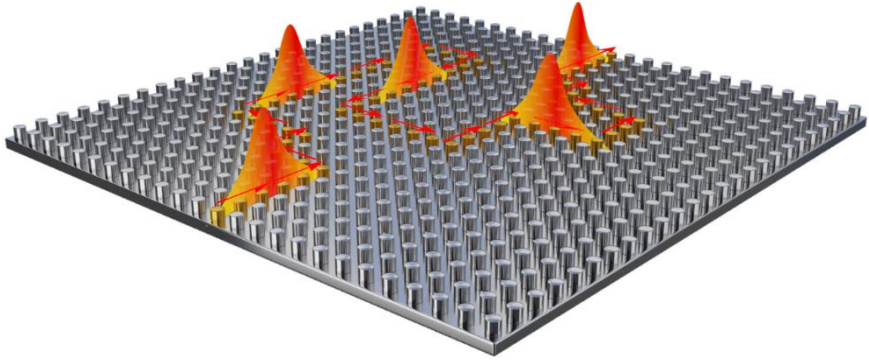

**Figure 1.** Schematic of a surface-wave pulse routing through multiple sharp corners along an arbitrary path in a surface-wave band-gap crystal.

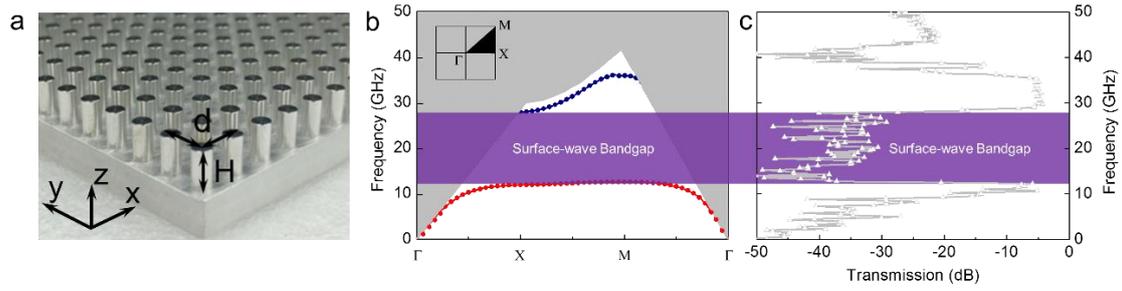

**Figure 2.** Experimental demonstration of a surface-wave band-gap crystal. a) Photograph of the surface-wave band-gap crystal. A square array of aluminium rods (with height $H = 5$ mm and periodicity $d = 5$ mm) stand on a metallic surface. b) Calculated band structure of the surface-wave band-gap crystal. Purple region represents the forbidden band gap. c) Measured transmission spectrum through the surface-wave band-gap crystal.

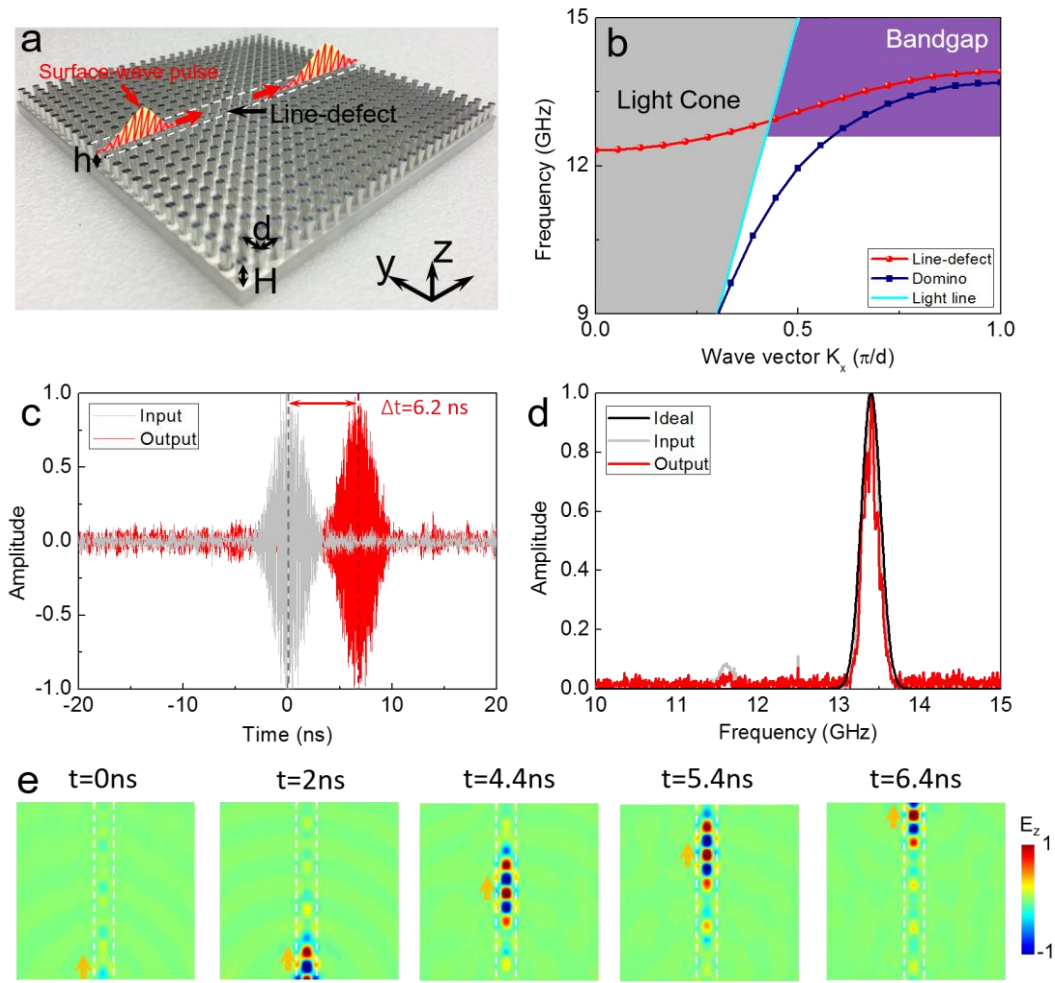

**Figure 3.** Surface-wave pulse propagation along a straight line-defect waveguide. a) Photograph of the straight line-defect waveguide formed by shortening a row of metallic rods from $H = 5$ mm to $h = 4.3$ mm in the middle of the surface-wave band-gap crystal. Inset shows the schematic representation of the pulse propagation. b) Dispersion of the line-defect waveguide compared to that of the traditional domino-plasmon waveguide. Purple region represents the forbidden band gap of the surface-wave band-gap crystal. c) Measured surface-wave pulse signals at the input and output of the waveguide. d) Spectral signatures of the measured input pulse and output pulse compared with the ideal case. e) Reconstructed surface-wave pulse propagation in the straight line-defect waveguide at different time frames.

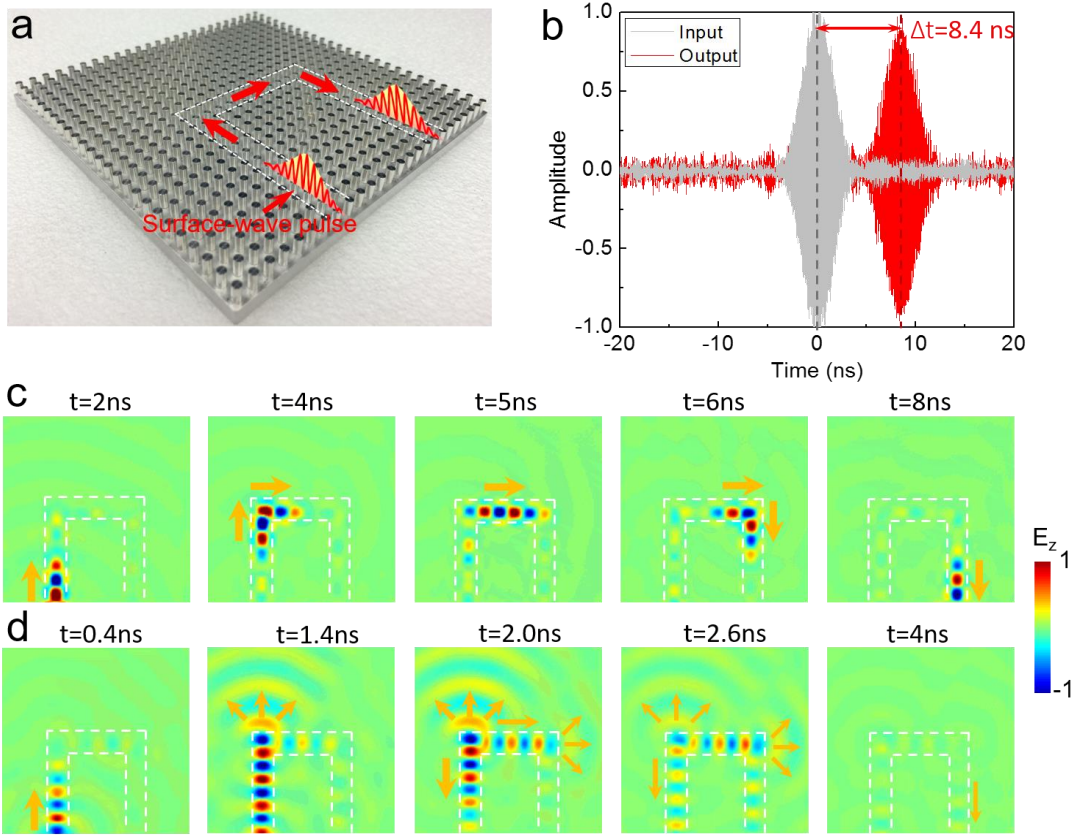

**Figure 4.** Surface-wave pulse propagation around a U-shaped line-defect waveguide with two sharp corners. a) Photograph of the U-shaped line-defect waveguide. Inset shows the schematic of the pulse propagation. b) Measured surface-wave pulse at the input and output of the U-shaped line-defect waveguide. c) Reconstructed surface-wave pulse propagation in the U-shaped line-defect waveguide at different time frames. d) Reconstructed surface-wave pulse propagation in the U-shaped domino-plasmon waveguide at different time frames.

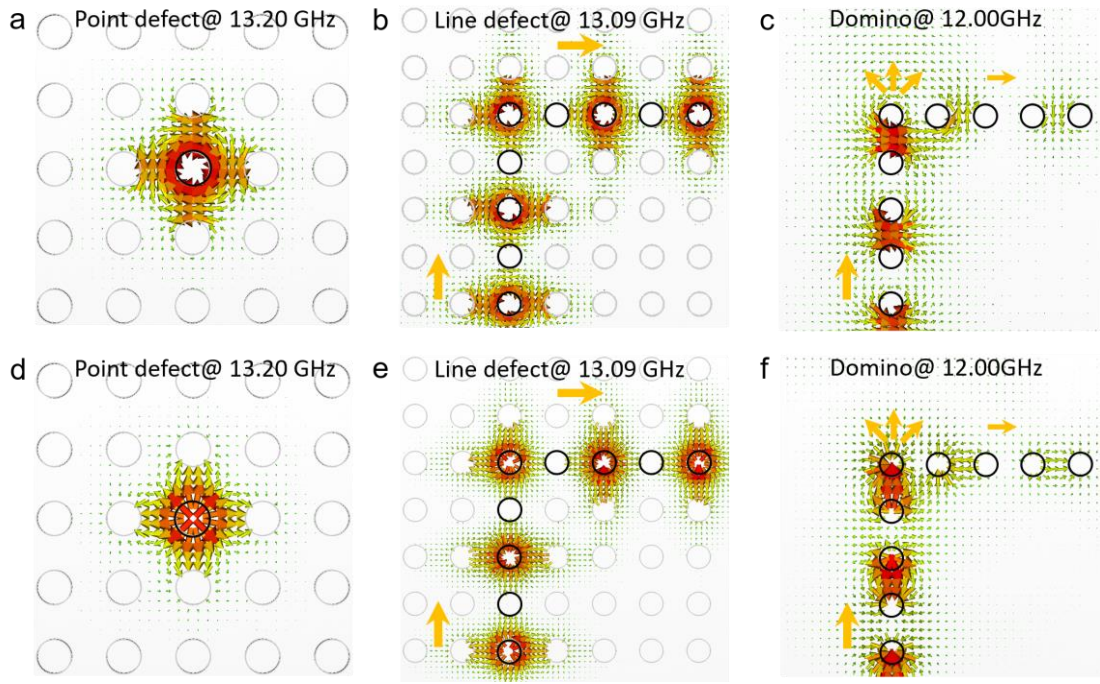

**Figure 5.** Field distributions of a point defect, a line-defect sharp bend in the surface-wave band-gap crystal and a domino-plasmon sharp bend without the surrounding surface-wave band-gap crystal. Magnetic (a-c) and electric (d-f) field distribution of (a,d) an isolated point defect at 13.20 GHz, (b,e) a line-defect sharp bend at 13.09 GHz and (c,f) a domino-plasmon sharp bend at 12.00 GHz in a transverse xy plane at half height of the 5-mm-high metallic rods. Black circles represent the line defect with shortened rods. Grey circles represent the surrounding surface-wave band-gap crystal.

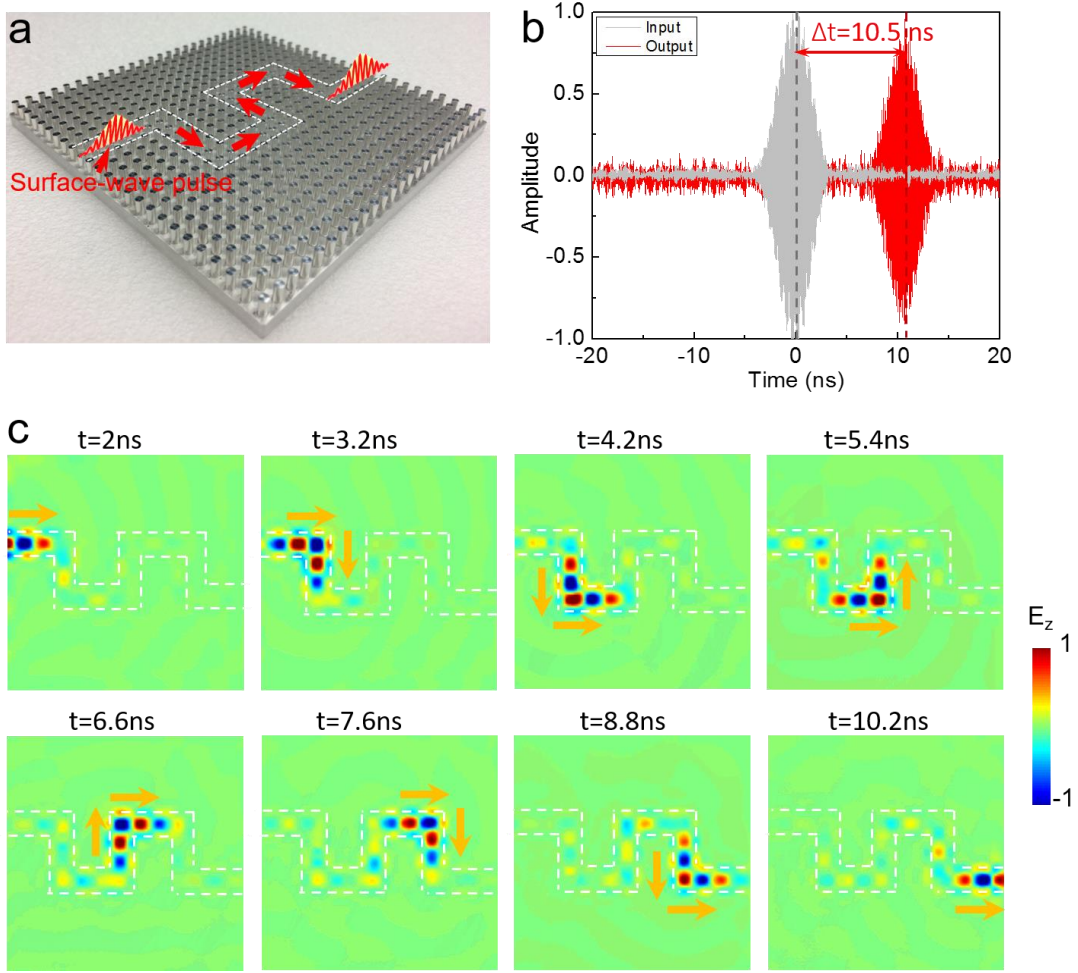

**Figure 6.** Surface-wave pulse propagation around a tortuous line-defect waveguide. a) Photograph of the tortuous line-defect waveguide. Inset shows the schematic of the pulse propagation. b) Measured surface-wave pulse signals at the input and output of the tortuous waveguide. c) Reconstructed surface-wave pulse propagation in the tortuous waveguide at different time frames.